\documentclass[aps,prl,reprint,superscriptaddress,floatfix]{revtex4-1}

\usepackage{amsmath}
\usepackage{amssymb}
\usepackage{graphicx}
\usepackage{bm}
\usepackage{xcolor}
\usepackage{url}
\usepackage[colorlinks]{hyperref}

\newcommand{\EN}{E_{\mathcal N}}
\newcommand{\Gee}{\Gamma_{\mathrm{ee}}}

\newcommand{\meff}{m_{\mathrm{eff}}}
\newcommand{\Tr}{\operatorname{Tr}}

\hypersetup{
plainpages=true,
breaklinks=true,
hypertexnames=false,
pageanchor=true,
colorlinks=true,
linkcolor=blue,
citecolor=blue,
urlcolor=blue
}

\begin{document}

\title{Quantum Wave-Particle Duality in Free-Electron--Free-Electron Entanglement}

\author{Du Ran}
\email{randu11111@163.com}
\affiliation{School of Electronic Information Engineering, Yangtze Normal University, Chongqing 408100, China}
\affiliation{Key Laboratory of Optical Chip and Intelligent Optoelectronic Systems, Chongqing Municipal Education Commission, Yangtze Normal University, Chongqing 408100, China}

\author{Jing-Yi Liu}
\affiliation{School of Electronic Information Engineering, Yangtze Normal University, Chongqing 408100, China}

\author{Dan Jin}
\affiliation{School of Electronic Information Engineering, Yangtze Normal University, Chongqing 408100, China}

\author{Peng-Yu Su}
\affiliation{School of Electronic Information Engineering, Yangtze Normal University, Chongqing 408100, China}
\affiliation{Key Laboratory of Optical Chip and Intelligent Optoelectronic Systems, Chongqing Municipal Education Commission, Yangtze Normal University, Chongqing 408100, China}

\author{Reuven Ianconescu}
\affiliation{Department of Electrical Engineering Physical Electronics, Center for Laser-Matter Interaction, Tel Aviv University, Ramat Aviv 69978, Israel}
\affiliation{Shenkar College of Engineering and Design 12, Anna Frank St., Ramat Gan, Israel}


\author{Shuai Liu}
\affiliation{School of Physics and Electronic Engineering, Hubei University of Arts and Science, Xiangyang 441053, China}

\author{Avraham Gover}
\email{gover@eng.tau.ac.il}
\affiliation{Department of Electrical Engineering Physical Electronics, Center for Laser-Matter Interaction, Tel Aviv University, Ramat Aviv 69978, Israel}

\date{\today}


\begin{abstract}
Point-particle descriptions of electron-electron interaction omit the coherent longitudinal extent of a free-electron quantum wave packet (QEW).
A relativistic two-electron wave-packet theory identifies the longitudinal QEW size as a direct control parameter for entanglement generated by mutual electromagnetic coupling.
For two electrons in spatially separated paths, the quadratic interaction phase gives a dimensionless entangling parameter $\Gee$ and the logarithmic negativity $\EN=\operatorname{arsinh}(\Gee)/\ln2$.
Full time-dependent Schr\"odinger equation calculations confirm the scaling for narrow QEWs and reveal higher-order Coulomb effects at larger spatial extent.
Free drift increases the interaction-point QEW width while preserving the momentum probability distribution, thereby enhancing the subsequently generated entanglement.
The results establish a direct connection between free-electron wave-particle duality and bipartite entanglement.
\end{abstract}

\maketitle

\textit{Introduction.}
Coherent control of free-electron quantum states has advanced through ultrafast electron microscopy \cite{zewail2010four,feist2017ultrafast,hassan2017high}, photon-induced near-field electron microscopy \cite{barwick2009photon,park2010photon,piazza2015simultaneous}, and optical or terahertz shaping of longitudinal, spectral, temporal, and transverse electron-wave-function degrees of freedom \cite{priebe2017attosecond,vanacore2018attosecond,reinhardt2020theory,madan2022ultrafast,chirita2022transverse,volkov2022spatiotemporal,yannai2023lossless,vanacore2020spatio,garcia2023spatiotemporal,tsesses2023tunable,velasco2025free,lavstovivckova2026quantum,wen2026nanoscale}.
A free electron prepared as a finite quantum electron wave packet (QEW) carries coherent spatial and spectral structures beyond the centroid variables of a point-particle description \cite{robicheaux2000coherent,robicheaux2000scattering,pan2019anomalous,feist2015quantum}.

A foundational finite-QEW analysis of stimulated radiative interaction established that the evolving longitudinal wave-packet size controls the transition between wave-like and point-particle-like electron response and retains the history of free propagation to the interaction region \cite{gover2018dimension}.
Subsequent studies extended QEW-size-dependent wave-particle behavior to interactions with radiation and bound quantum systems \cite{remez2019observing,wong2021control,gover2020free,zhang2021quantum,zhang2022quantum}, with particle-like dynamics following the electron centroid and wave-like dynamics depending on the coherent QEW phase structure \cite{garcia2021optical,ruimy2021toward}.
Engineered QEWs enable coherent excitation and scattering, quantum-state control, quantum radiation, interferometry, and quantum sensing \cite{yalunin2021tailored,morimoto2021coherent,ratzel2021controlling,zhang2025spontaneous,abad2025quantum,di2019probing,kfir2021optical,abad2024electron,ben2021shaping,karnieli2023quantum,bucher2023free,tsarev2021measurement}.
Quantum electron optics (QEO) has consequently expanded toward coherent electron-light interaction, recoil, nonclassical-light generation, strong coupling, and quantum-information protocols \cite{shiloh2022quantum,huang2023quantum,arend2025electrons,karnieli2024universal,velasco2024radiative,karnieli2023jaynes,karnieli2024strong,di2024toward,de2025roadmap}, while electron-electron interaction extends wave-packet physics to genuine bipartite quantum correlations \cite{schattschneider2020entanglement,ruimy2026signature}.

Electron-electron correlations arise from Coulomb repulsion, exchange, and collective many-electron dynamics \cite{talebi2021exchange,zandi2020transient}.
Very recent three-dimensional measurements of ultrashort two-electron pulses resolved joint energy--angle Coulomb correlations that are well reproduced by semiclassical point-particle simulations, while leaving the possible electron entanglement unquantified \cite{meier2026spatial}.
Genuine entanglement between identical electrons requires a physically defined subsystem partition, for which spatially separated propagation modes suppress ambiguity from overlap-induced exchange \cite{horodecki2009quantum,killoran2014extracting,friis2013fermionic}.
Two-electron microscopy has distinguished entanglement from classical Coulomb correlations and decoherence through correlation-sensitive state reconstruction \cite{tziperman2026two}.
More broadly, entanglement involving free electrons, light, and matter has become central to QEO, with electron-photon and related quantum correlations explored through cavity coupling, quantum erasure, coincidence detection, state tomography, and photonic-state engineering \cite{wang2020coherent,kfir2019entanglements,zhao2021quantum,baranes2022free,xie2025maximal,ruimy2025free,asban2021generation,gorlach2024photonic,henke2025probing,henke2025observation,preimesberger2025experimental,konevcna2022entangling,mechel2021quantum,rasmussen2024generation,baranes2023free,feist2022cavity,kazakevich2024spatial}.
Free-electron entanglement itself has been studied in electron-electron scattering \cite{schattschneider2020entanglement}, collective radiation \cite{karnieli2021superradiance}, nanophotonic mediation \cite{ruimy2026signature}, and heralded resource-transfer protocols \cite{ran2026heralded,ran2026heralded2}, while its direct dependence on the coherent longitudinal QEW size remains largely unexplored.

A point-particle description characterizes a free electron through its centroid trajectory and interaction phase, whereas a finite QEW coherently samples the interaction over its longitudinal extent \cite{pan2023weak,dahan2020resonant,echternkamp2016ramsey}.
For two interacting electrons, different longitudinal components acquire different pair phases, producing a spatially dependent nonlocal phase.
The present Letter develops a relativistic two-electron wave-packet theory for direct electron-electron entanglement and combines analytical treatment with full numerical wave-packet propagation to characterize wave-packet-size and higher-order Coulomb effects.
The resulting framework links the spatial structure of QEWs directly to the control and probing of free-electron entanglement.

\begin{figure}[t]
\centering
\includegraphics[width=0.44\textwidth]{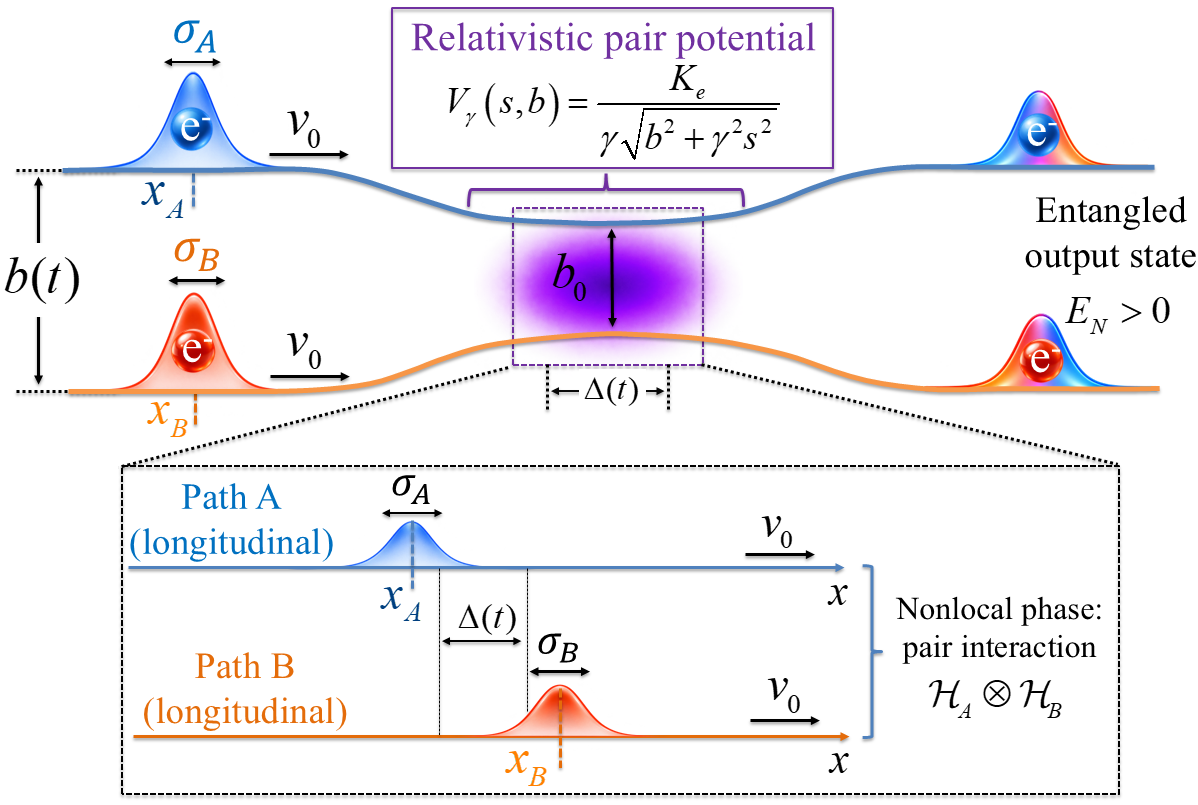}
\caption{
System model for free-electron--free-electron entanglement.
Two initially separable quantum electron wave packets propagate with mean velocity $v_0$ along spatially distinct paths $A$ and $B$.
Their transverse separation $b(t)$ reaches the minimum value $b_0$ near closest approach, while $\sigma_A$, $\sigma_B$, and $\Delta(t)$ denote the longitudinal wave-packet widths and centroid mismatch.
The enlarged interaction region shows the longitudinal degrees of freedom and the bipartition $\mathcal H_A\otimes\mathcal H_B$, where the accumulated nonlocal interaction phase generates an entangled output state.
}
\label{fig:system_model}
\end{figure}

\textit{Model and finite-wave-packet phase.}
Two free electrons propagate along spatially separated path modes $A$ and $B$, with one electron in each path, as shown in Fig.~\ref{fig:system_model}.
The two path modes define the operational bipartition $\mathcal H_A\otimes\mathcal H_B$, where $\mathcal H_j$ is the longitudinal Hilbert space associated with path $j$.
Each electron is described by a longitudinal QEW associated with its spatial channel.
For each path $j=A,B$, $\hat x_j$ denotes the longitudinal coordinate relative to the electron centroid and $\hat q_j=\hat p_j-p_0$ is the momentum deviation from the mean momentum $p_0=\gamma m_ev_0$.

The path geometry is characterized by the centroid separation $b(t)$ and the longitudinal centroid mismatch $\Delta(t)$.
Both electrons have the mean velocity $v_0=\beta c$, where $\beta=v_0/c$ and $c$ is the speed of light, with Lorentz factor $\gamma=(1-\beta^2)^{-1/2}$ and longitudinal effective mass $m_{\mathrm{eff}}=\gamma^3m_e$, where $m_e$ is the electron rest mass.
After removing the common energy and uniform longitudinal translation, the effective Hamiltonian is
\begin{align}
\hat H(t)
&=
\frac{\hat q_A^2+\hat q_B^2}{2\meff}
+V_\gamma[\Delta(t)+\hat x_A-\hat x_B,b(t)],
\nonumber\\
V_\gamma(s,b)
&=
\frac{K_e}{\gamma\sqrt{b^2+\gamma^2s^2}},
\qquad
K_e=\frac{e^2}{4\pi\epsilon_0}.
\label{eq:main_H}
\end{align}
Here $s$ denotes the longitudinal separation entering the pair kernel, $e>0$ is the elementary charge magnitude, and $\epsilon_0$ is the vacuum permittivity.
The interaction kernel $V_\gamma$ includes electric repulsion and magnetic reduction for co-propagating relativistic electrons.
Its longitudinal interaction scale is $b/\gamma$.
A derivation of the relativistic kernel and kinetic-energy expansion is given in Supplemental Material, Sec.~S1.

The transverse extent of each QEW is assumed to remain much smaller than the interpath separation $b(t)$ and approximately unchanged over the interaction region.
The transverse centroids therefore follow prescribed electron-optical paths, while transverse diffraction and finite-width averaging of the pair interaction are neglected within the fixed-path approximation.
Validity requires the interaction-induced transverse displacement to remain small compared with the path separation, together with negligible spin, radiative, and overlap-exchange effects.
Transverse overlap requires an antisymmetrized multimode description beyond the fixed-path model.
The trajectory functions $b(t)$ and $\Delta(t)$ determine the temporal variation of the electron-electron interaction.

At a longitudinal waist, each electron is described by a minimum-uncertainty Gaussian QEW, with $j=A,B$ labeling the two paths.
The wave function is
\[
\psi_j(x_j)=\frac{1}{(2\pi\sigma_j^2)^{1/4}}
\exp\left(-\frac{x_j^2}{4\sigma_j^2}\right),
\]
where $\sigma_j=\sqrt{\langle \hat{x}_j^2\rangle}$ is the rms longitudinal waist width and $\sigma_{q,j}=\hbar/(2\sigma_j)$ is the corresponding rms momentum width.
The separable two-electron input state is $\Psi_i(x_A,x_B)=\psi_A(x_A)\psi_B(x_B)$.
For weak longitudinal dispersion during the interaction, the outgoing state takes the phase-gate form
\[
\Psi_f(x_A,x_B)\simeq
\exp[i\Phi(x_A,x_B)]\Psi_i(x_A,x_B),
\]
where
$
\Phi(x_A,x_B)=
-\frac{1}{\hbar}\int dt\,
V_\gamma[\Delta(t)+x_A-x_B,b(t)]
$
is the accumulated electron-electron interaction phase.
Expanding $\Phi$ around the QEW centroids $x_A=x_B=0$ gives
\[
\Phi=\Phi_0+\phi_Ax_A+\phi_Bx_B
+\frac{\phi_{AA}}{2}x_A^2+\frac{\phi_{BB}}{2}x_B^2
+\chi_{AB}x_Ax_B+\mathcal O(x^3),
\]
where $\Phi_0$ is the global phase, $\phi_j$ and $\phi_{jj}$ are local phase coefficients, and $\chi_{AB}$ is the mixed quadratic coefficient.
Here $\mathcal O(x^3)$ denotes terms of total cubic and higher order in $x_A$ and $x_B$.
Global and local phase terms correspond to single-electron unitary operations.
At quadratic order, the mixed term $\chi_{AB}x_Ax_B$ produces the nonlocal phase responsible for entanglement.
The mixed coefficient and the corresponding dimensionless entangling parameter are
\begin{equation}
\chi_{AB}
=
\frac{1}{\hbar}
\int dt\,
\left.
\frac{\partial^2 V_\gamma(s,b(t))}{\partial s^2}
\right|_{s=\Delta(t)},
\
\Gee=2\sigma_A\sigma_B|\chi_{AB}|.
\label{eq:main_Gamma}
\end{equation}
The factor $\sigma_A\sigma_B$ measures the coherent longitudinal support available for resolving the interaction curvature.
In the formal point-particle limit $\sigma_A,\sigma_B\rightarrow0$, $\Gee$ approaches zero while the centroid Coulomb interaction remains finite.

After removal of the global and local phases, the quadratic interaction leaves the nonlocal factor $\exp(i\chi_{AB}x_Ax_B)$ in the two-electron Gaussian state.
Let $\hat\rho_{AB}$ denote the resulting two-electron density operator and $\hat\rho_A=\Tr_B(\hat\rho_{AB})$ the reduced state of electron $A$.
The reduced-state purity is
\[
\Tr(\hat\rho_A^2)=\frac{1}{\sqrt{1+\Gee^2}}.
\]
The geometric Schmidt spectrum gives the logarithmic negativity
\begin{equation}
\EN=
\log_2\left\|\hat\rho_{AB}^{T_B}\right\|_1
=
\frac{\operatorname{arsinh}\Gee}{\ln2},
\label{eq:main_EN}
\end{equation}
where $T_B$ denotes partial transposition with respect to electron $B$ and $\|\cdot\|_1$ is the trace norm.
For $\Gee\ll1$, $\EN\simeq\Gee/\ln2$.
The derivation and Gaussian validity conditions are given in Supplemental Material, Sec.~S2.

For the smooth trajectory, the transverse separation is taken as
$
b(t)=\sqrt{b_0^2+u_\perp^2t^2},
$
where $b_0$ is the minimum transverse separation and $u_\perp$ determines the separation rate.
For synchronous centroids $\Delta(t)=0$ and equal longitudinal widths
$\sigma_A=\sigma_B\equiv\sigma_x$, integration over $-\infty<t<\infty$ gives
\[
\Gamma_{\rm ee}
=
\frac{4K_e\gamma\sigma_x^2}
{\hbar u_\perp b_0^2}.
\]
The quadratic entangling strength therefore follows the scaling $\Gamma_{\rm ee}\propto\sigma_x^2$ and approaches zero as $\sigma_x\rightarrow0$.
The point-particle limit retains a finite centroid Coulomb interaction while the wave-packet-dependent longitudinal entangling contribution approaches zero.

\textit{Wave-packet-size control and quadratic scaling.}
For the full numerical calculation, the two-electron state evolves under the Hamiltonian in Eq.~\eqref{eq:main_H}, with the complete spatial dependence of $V_\gamma$ retained without a Taylor expansion.
The evolution is governed by the time-dependent Schr\"odinger equation (TDSE)
\begin{equation}
i\hbar\frac{\partial}{\partial t}
|\Psi(t)\rangle
=
\hat H(t)|\Psi(t)\rangle .
\label{eq:main_TDSE}
\end{equation}
For equal longitudinal masses, the center-of-mass and relative-coordinate operators are
$
\hat R= ({\hat x_A+\hat x_B})/{2},
\
\hat r=\hat x_A-\hat x_B.
$
In the corresponding coordinate representation, the Hamiltonian in Eq.~\eqref{eq:main_H} separates into
\[
\hat H_{\mathrm{CM}}
=-\frac{\hbar^2}{4m_{\mathrm{eff}}}\frac{\partial^2}{\partial R^2},
\
\hat H_{\mathrm{rel}}
=-\frac{\hbar^2}{2\mu}\frac{\partial^2}{\partial r^2}
+V_\gamma[\Delta(t)+r,b(t)],
\]
where $\mu=m_{\mathrm{eff}}/2$ is the reduced longitudinal mass.
Using the separation, Eq.~\eqref{eq:main_TDSE} is solved numerically in the center-of-mass and relative coordinates.
The center-of-mass component is propagated in momentum space, while the interacting relative-coordinate component is evolved with a second-order split-operator fast Fourier transform method.
The full two-electron wave function is then reconstructed in the $(x_A,x_B)$ coordinates and decomposed into Schmidt modes.
For normalized Schmidt amplitudes $s_n$ satisfying $\sum_n s_n^2=1$, the logarithmic negativity obtained from the TDSE is
\[
\EN^{\mathrm{TDSE}}
=
2\log_2\sum_n s_n .
\]

The calculations use a mean electron kinetic energy of $190~\mathrm{keV}$, consistent with recent two-electron microscopy experiments \cite{tziperman2026two}.
For the smooth trajectory, the characteristic interaction time is $\tau_b=b_0/u_\perp$, and numerical propagation covers $|t|\leq6\tau_b$.
Quadratic-potential and frozen-envelope calculations separately quantify longitudinal dispersion and higher-order spatial Coulomb contributions.
Numerical implementation and convergence tests are given in Supplemental Material, Sec.~S3.

\begin{figure}[t]
\centering
\includegraphics[width=0.47\textwidth]{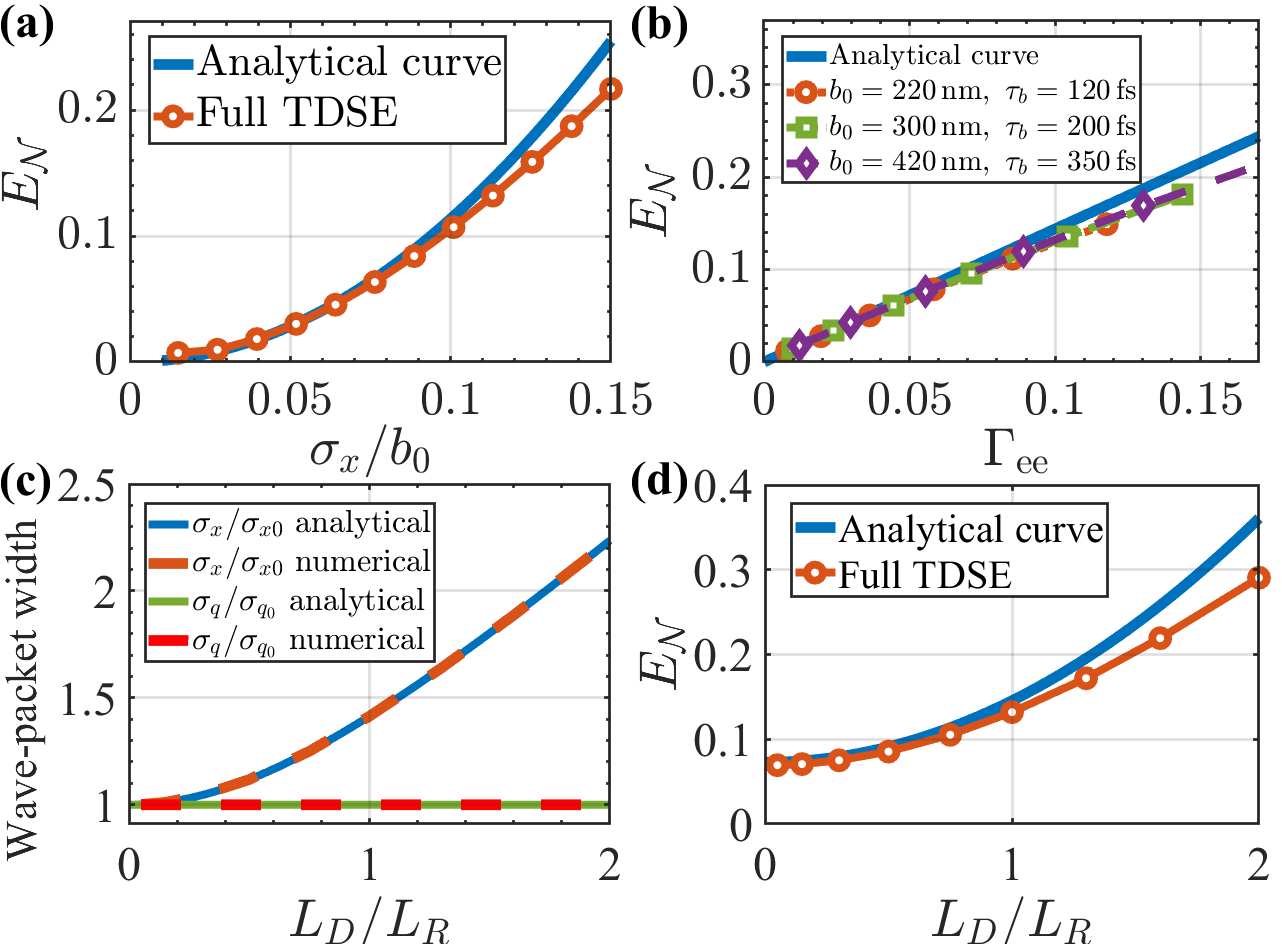}
\caption{
Wave-packet-size dependence, quadratic scaling, and drift control of free-electron--free-electron entanglement.
(a) Logarithmic negativity $\EN$ versus the normalized longitudinal width $\sigma_x/b_0$ for $b_0=300~\mathrm{nm}$ and $\tau_b=200~\mathrm{fs}$.
The solid line gives the quadratic analytical prediction, and the symbols show full TDSE results.
(b) Logarithmic negativity versus $\Gee$ for three interaction geometries.
The solid line denotes Eq.~\eqref{eq:main_EN}, and the symbols show the corresponding full TDSE results.
(c) Free longitudinal spreading of a QEW initially prepared with spatial width $\sigma_{x0}$ and momentum width $\sigma_{q0}$.
The spatial width increases with $L_D/L_R$, while the momentum width remains unchanged.
(d) Logarithmic negativity after drift-controlled wave-packet expansion.
All calculations use a mean electron kinetic energy of $190~\mathrm{keV}$.
}
\label{fig:size_scaling}
\end{figure}

Figure~\ref{fig:size_scaling}(a) shows the increase of $\EN$ with the coherent longitudinal width $\sigma_x$.
The quadratic analytical prediction agrees closely with the full TDSE calculation of Eq.~\eqref{eq:main_TDSE} for narrow QEWs.
Agreement confirms the leading scaling $\Gee\propto\sigma_x^2$ in the local spatial regime.
A gradual downward deviation appears as $\sigma_x/b_0$ increases, reflecting higher-order spatial contributions of the full interaction kernel.

Figure~\ref{fig:size_scaling}(b) tests the scaling with $\Gee$ for three interaction geometries.
Different values of $b_0$ and $\tau_b=b_0/u_\perp$ produce closely overlapping TDSE results when plotted against $\Gee$.
The result identifies $\Gee$ as the effective control parameter for quadratic entanglement generation.
Small deviations at larger $\Gee$ arise from higher-order spatial dependence of the relativistic interaction kernel.
Corresponding numerical checks and higher-order corrections are given in Supplemental Material, Sec.~S4.
The quadratic regime is governed by $\Gee$, while $\eta_{\mathrm{wp}}=\sqrt{2}\gamma\sigma_x/b_0$ characterizes the spatial range sampled by the QEW.
The analytical scaling applies for $\eta_{\mathrm{wp}}\ll1$ and weak longitudinal dispersion.
Increasing $\sigma_x/b_0$ enhances higher-order spatial contributions and produces the deviation from the quadratic prediction.

A direct change of the prepared Gaussian width also changes the minimum-uncertainty momentum width.
Free drift removes that ambiguity.
The drift-evolved QEW size was identified previously as the history-dependent variable governing the wave-particle transition in stimulated radiative interaction \cite{gover2018dimension}.
Here the same control principle is extended from a single-electron radiative response to a bipartite entanglement monotone.
For a QEW prepared at a longitudinal waist $\sigma_{x0}$, relativistic longitudinal dispersion gives
\begin{align}
\sigma_x(L_D)
&=
\sigma_{x0}
\sqrt{1+\left(\frac{L_D}{L_R}\right)^2},
\qquad
L_R=
\frac{2m_{\mathrm{eff}}v_{0}\sigma_{x0}^2}{\hbar},
\nonumber\\
\Gee(L_D)
&=
\Gee(0)
\left[
1+\left(\frac{L_D}{L_R}\right)^2
\right],
\label{eq:main_drift}
\end{align}
where $L_D$ is the free-drift length, $L_R$ is the relativistic longitudinal spreading length, and $\Gee(0)$ denotes the entangling parameter at the QEW waist.
Free propagation preserves the momentum probability distribution while increasing the spatial width.
The QEW also acquires a local quadratic chirp during the drift.

Figure~\ref{fig:size_scaling}(c) verifies the longitudinal spreading relation.
The normalized spatial width $\sigma_x/\sigma_{x0}$ increases with $L_D/L_R$, whereas the rms momentum width remains unchanged, with $\sigma_q(L_D)=\sigma_{q0}$ and $\sigma_{q0}=\hbar/(2\sigma_{x0})$ denoting the momentum width at the waist.
Figure~\ref{fig:size_scaling}(d) shows the corresponding increase of $\EN$ after the expanded QEWs enter the interaction region.
Every drift point originates from the same initial momentum probability distribution.
The variation of $\EN$ therefore directly demonstrates spatial wave-packet control of the electron-electron entanglement.
The analytical prediction remains accurate for moderate drift and gradually departs from the full TDSE result as higher spatial orders become relevant.
The drift protocol separates spatial wave-packet effects from changes in momentum bandwidth.
The centroid Coulomb interaction remains finite toward the point-particle limit, while the wave-packet-dependent entangling contribution decreases with the coherent longitudinal extent.
The drift-controlled variation of $\EN$ therefore reflects the spatial wave character of the QEW.

\textit{Quadratic node and cubic Coulomb entanglement.}
A longitudinal centroid mismatch provides direct access to higher-order spatial contributions of the electron-electron interaction.
For a constant mismatch $\Delta$ and the smooth trajectory defined above, subtraction of the irrelevant centroid phase yields the finite spatially dependent frozen-envelope phase
\begin{equation}
\Phi_C(s;\Delta)
=
\Lambda
\ln\!\left[
\frac{b_0^2+\gamma^2(\Delta+s)^2}
{b_0^2+\gamma^2\Delta^2}
\right],
\
\Lambda=\frac{K_e}{\gamma\hbar u_\perp}.
\label{eq:main_exact_phase}
\end{equation}
Here $s=x_A-x_B$ is the relative longitudinal coordinate and $\Lambda$ is the dimensionless interaction strength.
Introducing the normalized mismatch $\delta=\gamma\Delta/b_0$ gives
\begin{equation}
\Gee(\delta)
=
\Gee(0)
\frac{|1-\delta^2|}{(1+\delta^2)^2}.
\label{eq:main_mismatch}
\end{equation}
The quadratic contribution vanishes at $\delta_c=1$ for an infinite interaction trajectory.
For the numerical interval $|t|\leq6\tau_b$, the corresponding node shifts to $\delta_c=0.9752$.
The finite-window expression is given in Supplemental Material, Sec.~S5.
Figure~\ref{fig:cubic_node}(a) compares the quadratic prediction with the full TDSE calculation.
The quadratic entanglement decreases to zero at $\delta_c$, whereas the full Coulomb interaction retains a finite entanglement at the same mismatch.
A nonzero minimum also remains in the full TDSE curve near the quadratic node.
Quadratic-potential calculations reduce the residual entanglement to the numerical level, confirming the higher-order spatial origin of the finite signal.

\begin{figure}[t]
\centering
\includegraphics[width=0.46\textwidth]{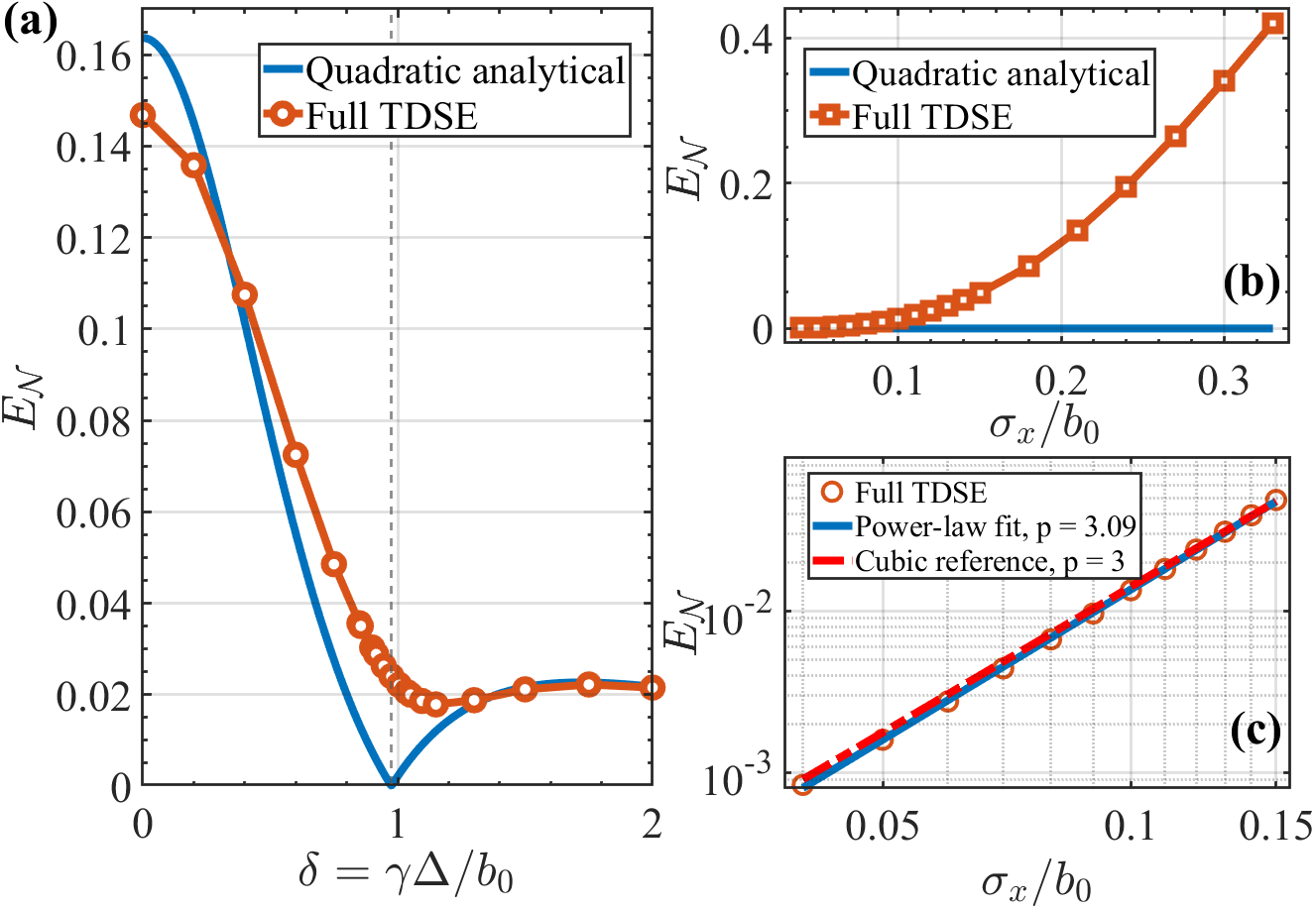}
\caption{
Quadratic entanglement node and higher-order Coulomb entanglement.
(a) Logarithmic negativity $\EN$ versus the normalized centroid mismatch $\delta=\gamma\Delta/b_0$ for $\sigma_x/b_0=0.12$.
The quadratic analytical result vanishes at the finite-window node marked by the vertical dashed line, while the full TDSE result remains finite.
(b) Wave-packet-size dependence at the quadratic node.
The quadratic contribution remains zero, whereas the full TDSE result increases with $\sigma_x/b_0$.
(c) Log-log dependence of the residual entanglement in the narrow-QEW regime.
The fitted exponent $p\simeq3.09$ agrees closely with the cubic reference $p=3$.
All calculations use $b_0=300~\mathrm{nm}$, $\tau_b=200~\mathrm{fs}$, and a mean electron kinetic energy of $190~\mathrm{keV}$.
}
\label{fig:cubic_node}
\end{figure}

The origin of the residual entanglement becomes transparent at the infinite-trajectory node $\delta_c=1$.
Defining $u=\gamma s/b_0$, Eq.~\eqref{eq:main_exact_phase} becomes
\begin{equation}
\Phi_C
=
\Lambda
\left(
u-\frac{u^3}{6}+\frac{u^4}{8}+\cdots
\right).
\label{eq:main_cubic}
\end{equation}
The linear contribution is local, while the quadratic contribution vanishes at the node.
The cubic term therefore provides the leading nonlocal contribution.
Expansion of $(x_A-x_B)^3$ contains the bipartite terms $x_A^2x_B$ and $x_Ax_B^2$.
The leading entangling contribution consequently scales as
$
\EN^{\mathrm{node}}
\propto
({\sigma_x}/{b_0})^3
$
in the narrow-QEW regime.
Figure~\ref{fig:cubic_node}(b) shows the wave-packet-size dependence of the residual entanglement at the quadratic node.
The full TDSE result approaches zero for narrow QEWs and increases rapidly with $\sigma_x/b_0$.
Figure~\ref{fig:cubic_node}(c) shows a fitted power-law exponent $p\simeq3.09$, in close agreement with the cubic exponent $p=3$ and confirming the cubic Coulomb entangling channel.

Recent advances in ultrafast electron microscopy provide the main ingredients required for the proposed scheme.
Few-electron photoemission has demonstrated strong pairwise Coulomb correlations and event-resolved detection at relativistic energies, while coincidence measurements have resolved individual electron-photon and recoil events \cite{haindl2023coulomb,yanagimoto2023time,preimesberger2025exploring}.
Coherent preparation, manipulation, and reconstruction of longitudinal free-electron wave functions are available with optically modulated electron beams \cite{feist2015quantum,priebe2017attosecond}, and electron interferometry provides beam-splitting and recombination elements for spatially separated coherent paths \cite{johnson2021scanning,johnson2022inelastic}.
A low-charge source suppresses uncontrolled many-electron interactions, while free propagation and relative arrival timing control the longitudinal QEW extent and centroid mismatch.
Optical modulation can therefore be confined to state preparation rather than to the direct electron-electron interaction itself.

Observation of the generated entanglement requires correlation-sensitive reconstruction beyond classical Coulomb observables.
Two-electron quantum-walk tomography has already distinguished point-particle, coherent matter-wave, classically correlated, and entangled regimes \cite{tziperman2026two}, while related photonic and photoelectron tomography demonstrates density-matrix reconstruction in free-electron and electronic systems \cite{gorlach2024photonic,laurell2025measuring}.
Such techniques provide a route to reconstructing the two-electron density matrix and evaluating the logarithmic negativity.
The main limitations arise from finite input purity, additional-electron interactions, transverse deflection, path overlap, and environmental decoherence.
Low-charge preparation and spatially separated paths suppress many-electron and exchange effects, while finite transverse width, transverse diffraction and deflection, path overlap, and decoherence require extensions beyond the present longitudinal pure-QEW model.

\textit{Conclusion.}
Direct electromagnetic coupling between two free electrons acquires wave-packet-dependent entangling power when the coherent longitudinal QEW structure is retained.
In the quadratic regime, the entanglement is governed by $\Gee$ and follows the closed relation $\EN=\operatorname{arsinh}(\Gee)/\ln2$.
Full Coulomb propagation confirms the wave-packet-size scaling, while free drift establishes spatial control at a fixed momentum probability distribution.
Suppression of the quadratic contribution further exposes a cubic Coulomb entangling channel governed by the longitudinal QEW size.
The results establish a direct connection between free-electron wave-particle duality and bipartite entanglement, providing spatial wave-packet control of quantum correlations between freely propagating electrons.

\begin{acknowledgments}
\textbf{Acknowledgments} The work is supported by the Natural Science Foundation of Chongqing (Grant No. CSTB2025NSCQ-GPX0416) and the Science and Technology Research Program of Chongqing Municipal Education Commission (Grant No. KJQN202401437).
P.-Y. Su acknowledges support from the Science and Technology Research Program of Chongqing Municipal Education Commission (Grant Nos. KJQN202401424 and KJQN202501434) and the Chongqing Fuling District Science and Technology Bureau (Grant No. FLKJ2026BAG1005).
S. Liu acknowledges the support of the Natural Science Foundation
of Hubei Province (Grant No. 2024AFB200), the Scientific Research Foundation of  the Education Department of Hubei Province (Grant No. B2025165), and the Research Start-up Fund of Hubei University of Arts and  Sciences (Grant No. qdf2022033).
A. G. and R. I. acknowledge the support of the Israel Science Foundation (Grant No. 2992124).
\end{acknowledgments}

\bibliographystyle{unsrt}
\bibliography{references}

\end{document}